\documentclass[graybox]{svmult}

\usepackage{type1cm}        
\usepackage{makeidx}         
\usepackage{graphicx}        
\usepackage{multicol}        
\usepackage[bottom]{footmisc}

\usepackage{newtxtext}       %
\usepackage[varvw]{newtxmath}       
\usepackage{amsmath}
\usepackage{amsthm}

\usepackage{latexsym}
\usepackage{amssymb}
\usepackage{bm}
\usepackage{bbm}
\usepackage{appendix}
\usepackage{graphics,epstopdf}
\usepackage{color}
\usepackage[colorlinks=true,linkcolor=blue,citecolor=red,plainpages=false,pdfpagelabels]{hyperref}
\usepackage[normalem]{ulem}
\usepackage{newlfont}
\usepackage{amsfonts}
\usepackage{amsthm}
\usepackage{graphicx}
\usepackage{epsfig}
\usepackage{amssymb,amsmath,graphicx,amsfonts,color,epsfig,amsxtra,amstext,latexsym,hyperref,setspace,amsthm,array,braket}
\usepackage[thinc]{esdiff}
\usepackage{float}
\usepackage{times}

\makeindex             

\begin{document}

\title*{Influence of gravity on the quantum speed limit in neutrino oscillations}
\author{Abhishek Kumar Jha\orcidID{0000-0003-3473-2464} \\ Mriganka Dutta\orcidID{0009-0000-3694-9240}\\ Subhashish Banerjee\orcidID{0000-0002-7739-4680} and\\ Banibrata Mukhopadhyay\orcidID{0000-0002-3020-9513}}

\institute{Abhishek Kumar Jha\at Department of Physics, Indian Institute of Science, Bangalore 560012, India, \email{kjabhishek@iisc.ac.in}
\and Mriganka Dutta \at Department of Physics, Indian Institute of Science, Bangalore 560012, India \email{mrigankad@iisc.ac.in}\and Subhashish Banerjee \at Indian Institute of Technology Jodhpur, Jodhpur 342011, India \email{subhashish@iitj.ac.in}\and Banibrata Mukhopadhyay \at Department of Physics, Indian Institute of Science, Bangalore 560012, India \email{bm@iisc.ac.in}}
\maketitle
\vspace{-1.5cm}
\textit{To be published in Astrophysics and Space Science Proceedings, titled ``The Relativistic Universe: From Classical to Quantum, Proceedings of the International Symposium on Recent Developments in Relativistic Astrophysics", Gangtok, December 11-13, 2023: to felicitate Prof. Banibrata Mukhopadhyay on his 50th Birth Anniversary", Editors: S Ghosh \& A R Rao, Springer Nature}
\vspace{1.2cm}

\abstract{The quantum speed limits (QSLs) determine the minimal amount of time required for a quantum system to evolve from an initial to a final state. We investigate QSLs for the unitary evolution of the neutrino-antineutrino system in the presence of a gravitational field. It is known that the transition probabilities between neutrino and antineutrino in the framework of one and two flavors depend on the strength of the gravitational field. The behavior of the QSL time in the two-flavor system indicates fast flavor transitions as the gravitational field strength increases. Subsequently, we observe quick suppression of entanglement by exploring the speed limit for entanglement entropy of two-flavor oscillations in the neutrino-antineutrino system in the proximity of a spinning primordial black hole.}
\abstract*{The quantum speed limits (QSLs) determine the minimal amount of time required for a quantum system to evolve from an initial to a final state. We investigate QSLs for the unitary evolution of the neutrino-antineutrino system in the presence of a gravitational field. It is known that the transition probabilities between neutrino and antineutrino in the framework of one and two flavors depend on the strength of the gravitational field. The behavior of the QSL time in the two-flavor system indicates fast flavor transitions as the gravitational field strength increases. Subsequently, we observe quick suppression of entanglement by exploring the speed limit for entanglement entropy of two-flavor oscillations in the neutrino-antineutrino system in the proximity of a spinning primordial black hole.}
\section{Introduction}
\label{Sec.1}
Several astrophysical and cosmological problems are involved with neutrino physics \cite{Balantekin:2013gqa}. Examples include neutrino-cooled accretion disks \cite{Chen:2006rra}, r-process nucleosynthesis in supernova explosions \cite{Surman:2008}, leptogenesis-baryogenesis \cite{Luty:1992un}, etc. On the other hand, neutrino oscillation is a quantum phenomenon where a neutrino, starting in one flavor state, can transit into a different flavor state as it progresses over time \cite{Pontecorvo:1957cp,Bilenky:2004xm}. This flavor transition occurs because neutrinos have non-zero mass. Any initial neutrino flavor states $\nu_\alpha$ (where, $\alpha=e,\mu,\tau$) can be represented as a linear superposition of three non-degenerate mass eigenstates. The study of neutrino oscillations in the combined two-flavor neutrino-antineutrino system has already been explored in the presence of gravity \cite{Sinha:2007uh}. 

In a locally flat coordinate system, gravitational interaction can be treated as an effective interaction \cite{Mukhopadhyay:2007vca}, when the neutrino Lagrangian density under gravity can be expressed as $\mathcal{L} = \mathcal{L}_f + \mathcal{L}_I$, where $\mathcal{L}_f$ corresponds to the free part, akin to the Lagrangian density in flat space, while the gravitational interaction part $\mathcal{L}_I$ results in an effective extension of the flat-space description leading to Lorentz invariance (LI), given by $\gamma^a\gamma^5B_a$. When $B_a$ remains constant within a locally flat coordinate system, LI is responsible for the charge-parity-time reversal (CPT) violation in gravity-induced neutrino-antineutrino systems \cite{Barger:1998xk,Barenboim:2002hx,Mohanty:2002et,Singh:2003sp,Ahluwalia:2004sz,Mukhopadhyay:2005gb}. This occurs in, e.g., the vicinity of rotating black holes (BH) \cite{Mukhopadhyay:2007vca,Singh:2003sp}, the anisotropic early universe phase \cite{Debnath:2005wk}.

In the Weyl representation, the term $\gamma^5$ in $\mathcal{L}_{I}$ leads to gravitational interactions represented as $\bar{\psi}\gamma^a\psi$ and $-\bar{\psi}^c\gamma^a\psi^c$ for neutrino and antineutrino, respectively. The superscript  “c" on $\psi$ indicates a charge-conjugate spinor or the spinor for the antiparticle. Consequently, the dispersion relations for neutrinos and antineutrinos differ and can be described as follows \cite{Sinha:2007uh,Mukhopadhyay:2007vca}:
\begin{equation}
    E_\psi=\sqrt{(\vec{p}-\vec{B})^2+m^2}+B_0, \hspace{0.5cm}
E_{\psi^c}=\sqrt{(\vec{p}+\vec{B})^2+m^2}-B_0,
\label{1}
\end{equation}
where $B_a$ is 4-vector gravitational potential and $\vec{p}$ the momentum of particles. Equation\,({\ref{1}}) effectively denotes gravitational Zeeman effect (GZE)
\cite{Barger:1998xk,Barenboim:2002hx,Mohanty:2002et,Singh:2003sp,Ahluwalia:2004sz, Mukhopadhyay:2005gb,Debnath:2005wk,Mukhopadhyay:2021ewd}. The detailed neutrino-antineutrino mixing and flavor oscillations in a two-flavor scenario were undertaken earlier \cite{Sinha:2007uh,Mukhopadhyay:2007vca}. The fundamental question arises: how quickly does the transition of the two-flavor neutrino-antineutrino states occur with the change of the gravitational field strength. In this work, we consider the unitary dynamics of two-flavor neutrino-antineutrino oscillations. We employ the quantum speed limit (QSL) time technique as a primary analytical tool to estimate the minimum time required for the evolution of two-flavor neutrino-antineutrino system.

The initial discovery of QSL stems from the uncertainty relationship between conjugate variables in quantum mechanics \cite{Mandelstam:1991,Margolus:1997ih}. The QSL time formulation depends on factors such as the shortest distance (or geodesic distance) between the initial and final states and the variance (or fluctuation) of the driving Hamiltonian \cite{Thakuria:2022taf}. Since, one of the most important aspects of quantum mechanics is the existence of superposition and entanglement, the deep understanding of the QSL has also led to the fundamental limitations on the rate of change of various entanglement measures \cite{Pandey:2022bph,Shrimali:2022bvt}. 
 
In recent times, entanglement in neutrino oscillations has been extensively discussed in various references \cite{Blasone:2007vw,Banerjee:2015mha,Alok:2014gya,KumarJha:2020pke,kumarJha: 2022,Jha:2022yik}. Additionally, quantification of entanglement measures in two-flavor neutrino-antineutrino oscillations within curved spacetime has also been done \cite{Dixit:2019lsg,Mukhopadhyay:2018oli}. Their results emphasize the existence of entanglement during the evolution of neutrino-antineutrino flavor states in the presence of gravity. In this work, we treat the two-flavor neutrino-antineutrino system as a bipartite quantum system. We delve into the behavior of entanglement entropy and the capacity of entanglement in two-flavor oscillations in the neutrino-antineutrino system, specifically near a spinning BH. We probe the QSL time for entanglement in the presence of low and high gravitational fields.

\section{Quantum speed limit for unitary quantum evolution and for entanglement}
\label{Sec.2}
  In the realm of quantum mechanics, the wave function $\ket{\psi}$ describes the state of a quantum system, while the dynamic governs its temporal evolution as 
  \begin{eqnarray}
      i\hbar\frac{d\ket{\psi}}{dt}=H\ket{\psi}
      \label{2}
\end{eqnarray}
  such that $\ket{\psi(t)}=e^{-i H t/\hbar}\ket{\psi}\equiv U_t\ket{\psi}$, where $U_t$ is the unitary time evolution operator, and the driving Hamiltonian $H$ is Hermitian and time independent. The QSL time, $T_{\text{QSL}}$, for such a system is defined as \cite{Thakuria:2022taf}
\begin{align}
T\geq T_{\text{QSL}},\hspace{0.5cm} T_{\text{QSL}}= \frac{\hbar S_0}{ \Delta {{H}}},
\label{3}
 \end{align}
where ${S_0} = \cos^{-1}(|\langle \psi (0)|\psi(T)\rangle|)=\arccos(\sqrt{P_s})$, is the geodesic distance between an initial and final states, and $P_s=|\langle \psi (0)|\psi(T)\rangle|^2$ is the survival probability of the initial state $\ket{\psi}$ at time $T$. A general representation of energy fluctuation $\Delta H$ when quantum system undergoes unitary dynamics is given by
\begin{equation}
   \Delta{H}\equiv \sqrt{ [\langle \dot{\psi} (t)|\dot{\psi}(t)\rangle-(i\langle \psi (t)|\dot{\psi}(t)\rangle)^2 ]},
    \label{4}
\end{equation}
where ${\ket{\dot{\psi}(t)}}\equiv d\ket{\psi(t)}/dt$. Equation\,(\ref{3}) suggests that $T_{\text{QSL}}$ is the minimum time required for the quantum system to evolve from $\ket{\psi(0)}$ to $\ket{\psi(T)}$ by unitary evolution.

A bipartite quantum system, in the context of quantum mechanics, is a composite quantum system that consists of two individual subsystems or components \cite{Nielsen:2012yss}. Each of these subsystems can be thought of as a separate quantum system, and they are often described by their own quantum states and associated Hilbert spaces. For unitary quantum evolution, the quantum state of the entire bipartite system is typically described using a mathematical framework called a density matrix: $\rho=\ket{\psi}\bra{\psi}$ and it is often represented as a tensor product of the quantum states of the individual subsystems: $\rho=\rho_A \otimes \rho_B \in\mathcal{H}_A\otimes\mathcal{H}_B$. It is worth mentioning that due to the unitary evolution, $\rho$ always exists in a pure quantum state with properties such as $\rho=\rho^2$ (idempotent matrix) and $\mathrm{Tr}(\rho^2)=1$. The reduced density matrix $\rho_{A}= \mathrm{Tr}_B(\rho)$ and $\rho_B=\mathrm{Tr}_A(\rho)$ follows $ \mathrm{Tr}(\rho_A^2)<1$ and $\mathrm{Tr}(\rho_B^2)<1$, which are in the mixed states. An example of a bipartite pure quantum system is Bell's state. Furthermore, $\rho$ accounts for the entanglement and correlations that can exist between the two subsystems. Within a bipartite pure quantum system, QSL for generating entanglement is intricately tied not only to fluctuations in the Hamiltonian but also inversely linked to the time-averaged square root of the capacity of entanglement. In scenarios where the driving Hamiltonian remains time-independent, the quantum speed limit for entanglement adheres to the following bound \cite{Shrimali:2022bvt}
\begin{equation}
    T\geq T^E_{\text{QSL}}=\frac{\hbar|S_{EE}(T)-S_{EE}(0)|}{2\Delta H \frac{1}{T}\int ^T_0 \sqrt{C_E(t)}dt},
    \label{5}
\end{equation}
where the entanglement entropy $S_{EE}$ is defined as
\begin{equation}
   S_{EE}=S(\rho_A)=-\mathrm{Tr}(\rho_A log_2\rho_A),
    \label{6}
\end{equation}
for the reduced density matrix $\rho_A$. The variance (or fluctuation) in the entanglement entropy operator $S_{EE}$ is defined as capacity of entanglement $C_{E}$, given as
\begin{equation}
    C_E=\sum_i\lambda_i log_2^2\lambda_i-S^2_{EE},
    \label{7}
\end{equation}
where $\lambda_i's$ are the eigenvalue of $\rho_A$ (or $\rho_B$). Equation\,(\ref{5}) establishes a QSL on the production and decay of entanglement in the time-evolved state $\ket{\psi(t)}$.

\section{Four-vector gravitational potential around a rotating compact object}
\label{Sec.3}
We consider a rotating BH described by the Kerr geometry. We derive the analytical expression of the four-vector gravitational potential for a rotating BH, taking into account Kerr geometry in the Kerr-Schild form \cite{Doran:1999gb}. Our variables are denoted by $t~(= x_0)$, $x~(= x_1)$, $y~(= x_2)$, and $z~(= x_3)$. In this Cartesian form, the Kerr metric with signature [+ - - -] can be written as
\begin{equation}
ds^2=\eta_{ij} dx^{i} dx^{j} - \left[ \frac{2\alpha}{\rho} s_i v_j +\alpha^2 v_i v_j\right] dx^i dx^j,
\label{8}
\end{equation}
where
\begin{equation}
\alpha=\frac{\sqrt{2Mr}}{\rho},\hspace{1cm} \rho^2=r^2+\frac{a^2 z^2}{r^2},
\label{9}
\end{equation}

\begin{equation}
s_i=\left( 0, \frac{rx}{\sqrt{r^2+a^2}}, \frac{ry}{\sqrt{r^2+a^2}}, \frac{z\sqrt{r^2+a^2}}{r} \right), \hspace{0.5cm} v_i=\left( 1, \frac{ay}{a^2+r^2}, -\frac{ax}{a^2+r^2}, 0 \right).
\label{10}
\end{equation}
Here $a$ and $M$ are respectively the specific angular momentum and mass of the Kerr BH and $r$ is a positive definite coordinate satisfying the following equation,
\begin{equation}
r^4-r^2(x^2+y^2+z^2-a^2)-a^2 z^2=0.
\label{11}
\end{equation}
Here we choose $G=M=c=\hbar=1$. The four-vector $B^d$ can be calculated by \cite{Mukhopadhyay:2005gb}

\begin{equation}
B^d=\epsilon^{abcd}e_{b\lambda}\left(\partial_a e_c^{\lambda} + \Gamma_{\alpha\mu}^{\lambda} e_c^{\alpha}e_a^{\mu}\right),
\label{12}
\end{equation}
which leads to
\begin{equation}
    B^0=-\frac{2\sqrt{2M}az}{\left[\rho^2 \sqrt{r(a^2 + r^2)}\right]},
    \label{13}
\end{equation}
\begin{align}
    B^1 &=a\alpha^2xz\left[\frac{a^2\alpha^2(x^2+y^2)+(\alpha^2+3)(a^2+r^2)^2}{\rho^2(a^2+r^2)^3}\right],\nonumber\\
     B^2 &=-a\alpha^2yz\left[\frac{a^2\alpha^2(x^2+y^2)+(\alpha^2+3)(a^2+r^2)^2}{\rho^2(a^2+r^2)^3}\right],\nonumber\\
     B^3 &=-\frac{a\alpha^2r^2}{\rho^2(a^2+r^2)^4}\left[(a^2+r^2)^2(x^2+y^2)(3+\alpha^2)+a^2\alpha^2(x^2+y^2)^2\right].
    \label{14}
\end{align}

\section{Quantum speed limit for the evolution of two-flavor neutrino-antineutrino system in the presence of gravity}
\label{Sec.4}

\begin{figure}[t]
\sidecaption
\includegraphics[scale=0.69]{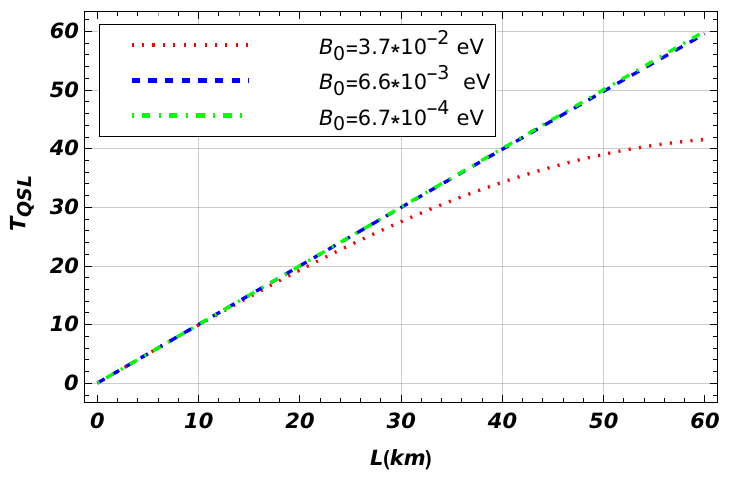}
    \caption{In the two-flavor scenario, ${T}_{\text{QSL}}$ of the initial state $\ket{\psi_e}$ as a function of propagation length ${L\text{(km)}}$ at three distinct values of the gravitational potential, i.e. at three different distance from BH, given in Table\,\ref{T1}. The other parameters used are: $m_e=5\times 10^{-3}\text{eV}$, $m_\mu=6.5\times 10^{-3}{\text{eV}}$, $m_{e\mu}=3.5\times 10^{-3}\text{eV}$.}
    \label{Fig.1}
   \end{figure}
   
We strictly follow earlier results \cite{Sinha:2007uh} to obtain QSL for a evolved neutrino-antineutrino flavor state under gravity. The evolving flavor states $\ket{\psi_e^c(t)}$, $\ket{\psi_\mu^c(t)}$, $\ket{\psi_e(t)}$ and $\ket{\psi_\mu(t)}$ can also be projected to flavor basis in the form 
\begin{equation}
     \begin{pmatrix}
       \ket{\psi_e^c(t)}\\
        \ket{\psi_\mu^c(t)}\\
        \ket{\psi_e(t)}\\
        \ket{\psi_\mu(t)}\\
    \end{pmatrix}=\mathbf{T}^{-1}\text{diag}(e^{-iE_1 t},e^{-iE_2 t},e^{-iE_3 t},e^{-iE_4 t}) \mathbf{T}\begin{pmatrix}
       \ket{\psi_e^c}\\
        \ket{\psi_\mu^c}\\
        \ket{\psi_e}\\
        \ket{\psi_\mu}\\
    \end{pmatrix},
    \label{15}
\end{equation}
where $\ket{\psi_e^c}$, $\ket{\psi_\mu^c}$, $\ket{\psi_e}$ and $\ket{\psi_\mu}$ are flavor states at time $t=0$. Thus, using Eq.\,(\ref{15}), the time evolved electron flavor neutrino state in superposition of flavor basis can be obtained as
\begin{equation}
   \ket{\psi_e(t)}=T_{e e^c}(t)\ket{\psi^c_e} +T_{e \mu^c}(t)\ket{\psi^c_\mu}
   + T_{ee}(t) \ket{\psi_e}+T_{e \mu}(t) \ket{\psi_\mu}.
   \label{16}
\end{equation}

The mixing angles for neutrino-antineutrino and electron-muon neutrino mixing are related to the masses and the gravitational scalar potential,
respectively, given by \cite{Sinha:2007uh}
\begin{equation}
\tan\theta_{e,\mu}=\frac{m_{e,\mu}}{B_0+\sqrt{(B_0)^2+m^2_{e,\mu}}},
    \label{17}
\end{equation}
\begin{equation}
    \tan\phi_{1,2}=\frac{\mp 2 m_{e\mu}}{m_{e(1,2)}-m_{\mu(1,2)}+\sqrt{(m_{e(1,2)}-m_{\mu(1,2)})^2+4m^2_{e\mu}}}.
    \label{18}
\end{equation}
The masses corresponding to the mass eigenstates are given as
\begin{eqnarray}
    {M_{1,2}=\frac{1}{2}[(m_{e1}+m_{\mu 1})\pm\sqrt{(m_{e1}-m_{\mu 1})^2+4m^2_{e\mu}}],}&\nonumber\\
    {M_{3,4}=\frac{1}{2}[(m_{e2}+m_{\mu 2})\pm\sqrt{(m_{e2}-m_{\mu 2})^2+4m^2_{e\mu}}],}&
    \label{19}
\end{eqnarray}
with
\begin{equation}
    m_{(e,\mu)1}=-\sqrt{(B_0)^2+m^2_{e,\mu}}, \hspace{0.5cm} m_{(e,\mu)2} =\sqrt{(B_0)^2+m^2_{e,\mu}}.
    \label{20}
\end{equation}
Assuming neutrinos to be ultra-relativistic particles, energies are
\begin{equation}
   E_1=E_2=|\vec{p}|+|\vec{B}|- B_0, \hspace{0.5cm}
   E_3=E_4=|\vec{p}|-|\vec{B}|+ B_0.
   \label{21}
\end{equation}

 Thus, we find the survival probability of the initial electron flavor neutrino state $\ket{\psi_e}$ using Eq.\,(\ref{16}) as 
\begin{equation}
    P_s=|T_{ee}(t)|^2.
    \label{22}
\end{equation}

 \begin{table}[!t]
\caption{Three different gravitational scalar-potential, $B_0(\text{eV})$, and vector potential magnitude, $|\vec{B}|(\text{eV})$, corresponding to three different radii, $100$, $200$ and $500$, from the BH.}
\label{T1}  
\begin{tabular}{p{3.8cm}p{3.5cm}p{3.9cm}}
\hline\noalign{\smallskip}
$r$ & $B_0 (\text{eV})$ & $|\vec{B}|(\text{eV})$\\
\noalign{\smallskip}\svhline\noalign{\smallskip}
100 & $3.7\times 10^{-2} $ & $1.3\times10^{-2}$\\
200 & $6.6\times10^{-3}$ & $1.7\times10^{-3}$\\
500 &  $6.7\times10^{-4}$ & $1.1\times10^{-4}$\\
\noalign{\smallskip}\hline\noalign{\smallskip}
 \end{tabular}
 \end{table}

 Furthermore, using Eq.\,(\ref{4}) and Eq.\,(\ref{16}), we compute energy fluctuation $\Delta H$. Subsequently, using Eq.\,(\ref{3}) and Eq.\,(\ref{22}), we estimate $T_{\text{QSL}}$ for the initial electron flavor neutrino state $\ket{\psi_{e}}$ as
\begin{equation}
    T_{\text{QSL}}=\frac{\arccos(\sqrt{|T_{ee}(t)|^2})}{\Delta{H}}.
    \label{23}
\end{equation}
 
We assume that neutrinos are moving around a rotating BH at fixed radii. In the ultra-relativistic limit, $t\approx L$, when $L$ is the propagation length. Fig.\,\ref{Fig.1} illustrates $T_{\text{QSL}}$ of the initial electron flavor neutrino state $\ket{\psi_e}$ as a function of propagation length $L$ for different values of the gravitational scalar potential, as mentioned in Table\,\ref{1}. We observe that the behavior of $T_{\text{QSL}}$ vs. $L$ at low $B_0$ (blue dashed and green dot-dashed lines) is tight, which means $T_{\text{QSL}}/L$ equals 1. Therefore, the evolutionary speed of the initial state $\ket{\psi_e}$ is unaltered and the minimum time taken to oscillate is an extremum. However, as the $B_0$ changes from a smaller to a higher value (red dotted line), the initial neutrino state $\ket{\psi_e}$ begins to oscillate faster because the ratio $T_{\text{QSL}}/L<1$, causing the dynamical evolution to speed up as a smaller value of $T_{\text{QSL}}/L$ implies a faster evolution of the quantum state. Hence, it is evident from Eq.\,(\ref{23}) that the QSL time of two-flavor neutrino-antineutrino oscillations changes significantly depending on the gravitational field strength.

\section{Quantum speed limit for entanglement in two-flavor neutrino-antineutrino oscillations in the presence of gravity}
\label{Sec.5}

\begin{figure}[t]
\sidecaption
\includegraphics[scale=0.69]{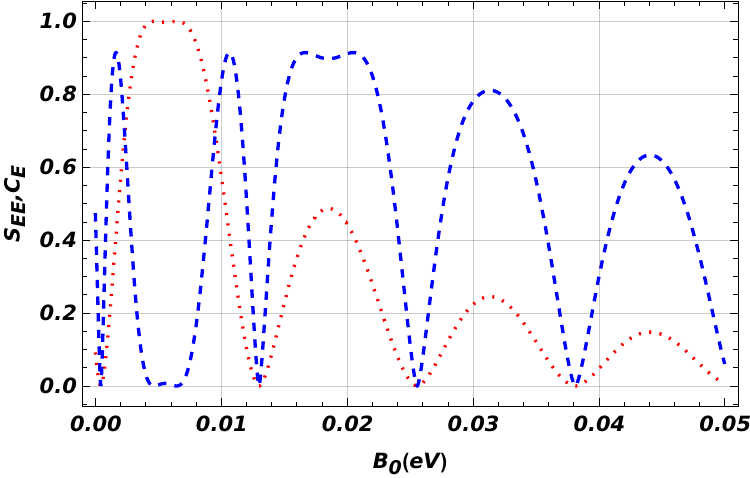}
    \caption{In the two-flavor scenario, the entanglement entropy, $S_{EE}$ (red dotted line), and the capacity of entanglement, $C_E$ (blue dashed line), of the initial state $\ket{\psi_e}$ as functions of gravitational scalar potential $B_0$. The other parameters are: $m_e=5\times 10^{-3}\text{eV}$, $m_\mu=6.5\times 10^{-3}{\text{eV}}$, $m_{e\mu}=3.5\times 10^{-3}\text{eV}$ and $L=300\text{km}$.}
    \label{Fig.2}
    \end{figure}

Now, we treat the system of the two-flavor neutrino-antineutrino system as a four-qubit system under the impact of neutrino-antineutrino mixing caused by the gravitational field. The representation of the occupation number can be expressed as \cite{Dixit:2019lsg}
\begin{eqnarray}
   { \ket{\psi_e^c}\equiv \ket{1}_{e^c}\otimes\ket{0}_{\mu^c}\otimes \ket{0}_{e}\otimes\ket{0}_\mu\equiv\ket{1000}}&\nonumber,\\
   { \ket{\psi_\mu^c}\equiv \ket{0}_{e^c}\otimes\ket{1}_{\mu^c}\otimes \ket{0}_{e}\otimes\ket{0}_\mu\equiv\ket{0100}}&\nonumber,\\
   { \ket{\psi_e}\equiv \ket{0}_{e^c}\otimes\ket{0}_{\mu^c}\otimes \ket{1}_{e}\otimes\ket{0}_\mu\equiv\ket{0010}}&\nonumber,\\
   { \ket{\psi_\mu}\equiv \ket{0}_{e^c}\otimes\ket{0}_{\mu^c}\otimes \ket{0}_{e}\otimes\ket{1}_\mu\equiv\ket{0001}}&\nonumber.\\
   \label{24}
\end{eqnarray}

Thus, using Eq.\,(\ref{16}), the time evolution of electron flavor neutrino state in four-qubit system can be written as
  \begin{equation}
   \ket{\psi_e(t)}=T_{e e^c}(t)\ket{1000}+T_{e \mu^c}(t)\ket{0100}
   + T_{ee}(t) \ket{0010}+T_{e \mu}(t) \ket{0001}.
   \label{25}
\end{equation}  

Furthermore, using Eq.\,(\ref{25}), we construct a $16\times16$ density matrix, $\rho(t)=\ket{\psi_e(t)}\bra{\psi_{e}(t)}$. Here, $\rho^2(t)=\rho(t)$ and $\mathrm{Tr}(\rho^2(t))=1$, which means that the $\rho(t)$ is in a pure state. We look for a bipartite scenario where the reduced density matrix is obtained by tracing $\rho(t)$ over the other two qubits. The required $4\times4$ reduced density matrix can be expressed as
\begin{equation}
 \rho_{\text{red}}(t)=\text{diag}(0,0,\lambda_1,\lambda_2),
 \label{26}
\end{equation}
where $\rho^2_{\text{red}}(t)\neq\rho_{\text{red}}(t)$ and $\mathrm{Tr}(\rho_{\text{red}}^2(t))<1$, signifying $\rho_{\text{red}}(t)$ is in a mixed state. Consequently, the state $\ket{\psi_e(t)}$ is a bipartite pure state. Here, $\lambda_1$ and $\lambda_2$ represent the two non-zero eigenvalues of $\rho_{\text{red}}(t)$.
Subsequently, putting $\lambda_1$ and $\lambda_2$ values in Eqs.\,(\ref{6}) and (\ref{7}), we calculate entanglement entropy 
\begin{equation}
    S_{EE}=-\lambda_1 log_2\lambda_1-\lambda_2 log_2\lambda_2,
    \label{27}
\end{equation}
 and capacity of entanglement
\begin{equation}
    C_E=(\lambda_1 log_2^2\lambda_1+\lambda_2 log_2^2\lambda_2)-(-\lambda_1 log_2\lambda_1-\lambda_2 log_2\lambda_2)^2,
    \label{28}
\end{equation}
of the state $\ket{\psi_e(t)}$. 

\begin{figure}[t]
\sidecaption
\includegraphics[scale=.69]{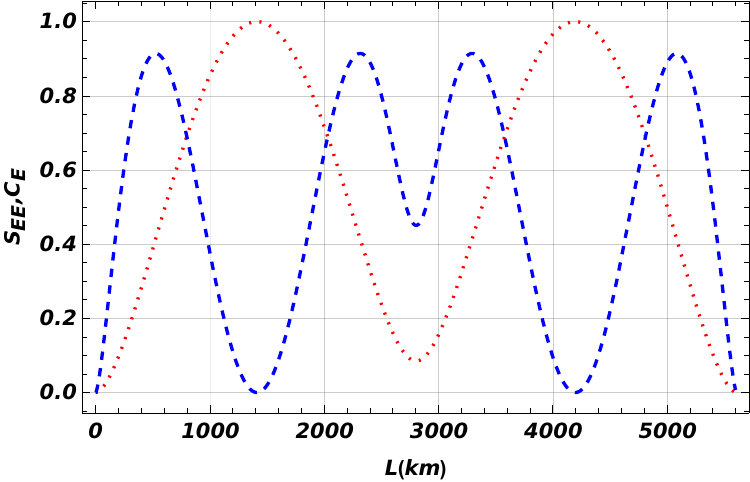}
   \caption{In the two-flavor scenario, the entanglement entropy, $S_{EE}$ (red dotted line), and the capacity of entanglement, $C_E$ (blue dashed line), of the initial state $\ket{\psi_e}$ as functions of propagation length $L\text{(km)}$ at low value of gravitational scalar potential $B_0=6.7\times10^{-4} \text{eV}$. The other parameters are: $m_e=5\times 10^{-3}\text{eV}$, $m_\mu=6.5\times 10^{-3}{\text{eV}}$, $m_{e\mu}=3.5\times 10^{-3}\text{eV}$.}
    \label{Fig.3}
   \end{figure}
\begin{figure}[t]
\sidecaption
\includegraphics[scale=0.69]{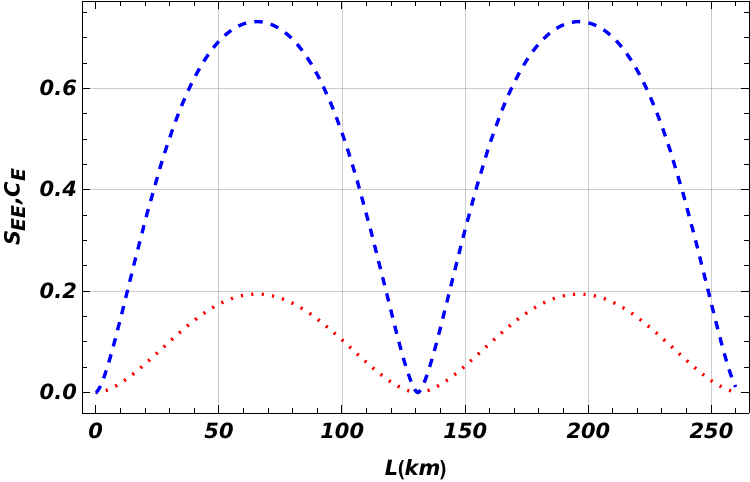}
   \caption{In the two-flavor scenario, the entanglement entropy, $S_{EE}$ (red dotted line), and the capacity of entanglement, $C_E$ (blue dashed line), of the initial state $\ket{\psi_e}$ as functions of propagation length $L\text{(km)}$ at high value of gravitational scalar potential $B_0=3.7\times 10^{-2}\text{eV}$. The other parameters are: $m_e=5\times 10^{-3}\text{eV}$, $m_\mu=6.5\times 10^{-3}{\text{eV}}$, $m_{e\mu}=3.5\times 10^{-3}\text{eV}$.}
    \label{Fig.4}
   \end{figure}

In the ultra-relativistic limit, $t\approx L$. In Fig.\,\ref{Fig.2}, at a fixed propagation length of $L=300\text{km}$, we depict the variation of $S_{EE}$ (red dotted line) and $C_E$ (blue dashed line) for the initial state $\ket{\psi_e}$ as functions of gravitational scalar potential $B_0$. We observe a decrease in entanglement entropy ($S_{EE}$) as $B_0$ increases. In Figs.\,\ref{Fig.3} and \ref{Fig.4}, we estimate $S_{EE}$ (red dotted line) and $C_{E}$ (blue dashed line) as functions of propagation length $L$ for the initial state $\ket{\psi_e}$ at low and high values of $B_0$. 
At $L=0$, $\lambda_1$ and $\lambda_2$ tend to 0 and 1, respectively. This implies that at $L=0$, both $S_{EE}$ and $C_E$ vanish. At $L>0$, $S_{EE}>0$ and $C_{E}>0$. In Fig.\,\ref{Fig.3}, we observe that at a low value of $B_0=6.7\times 10^{-4}\text{eV}$, $S_{EE}$ (red dotted line) develops monotonically towards the maximum. However, $C_E$ (blue dashed line) goes to zero in both the maximally entangled and separable state limits, while reaching a maximum in a partially entangled state with varying $L$. This result shows that the bipartite entanglement persists during the time evolution of the initial electron flavor neutrino state $\ket{\psi_e}$. However, the comparison between Fig.\,\ref{Fig.3} and Fig.\,\ref{Fig.4} reveals that as we vary the strength of $B_0$ from a low ($B_0=6.7\times 10^{-4}\text{eV}$) to a high value ($B_0=3.7\times 10^{-2}\text{eV}$), entanglement in the state $\ket{\psi_e(t)}$ gets suppressed.

 \begin{figure}[t]
\sidecaption
\includegraphics[scale=.69]{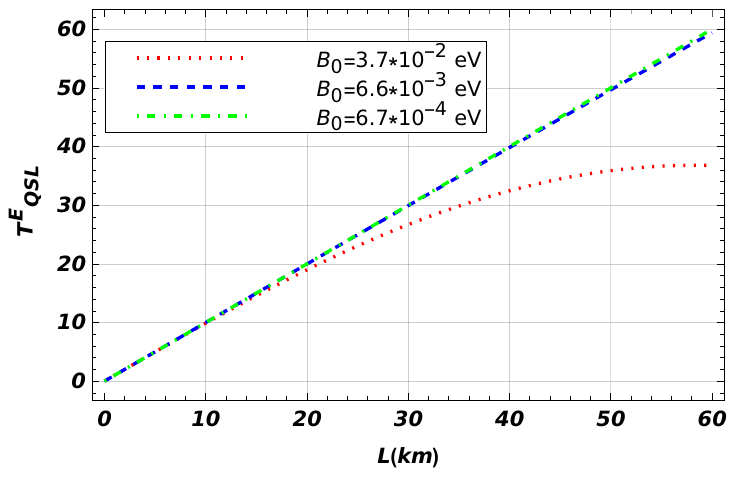}
   \caption{Same as Fig.\,\ref{Fig.1}, except $T^E_{\text{QSL}}$.}
    \label{Fig.5}
   \end{figure}

Further, using Eqs.\,(\ref{27}) and (\ref{28}) in Eq. (\ref{5}), QSL for entanglement can be obtained for the state $\ket{\psi_e(t)}$ as
\begin{equation}
    T^E_{\text{QSL}}=\frac{| -\lambda_1 log_2\lambda_1-\lambda_2 log_2\lambda_2|}{\Delta{H}\frac{1}{T}\int_0^T\sqrt{C_E(t)}dt}.
    \label{29}
\end{equation}

In Fig.\,\ref{Fig.5}, we depict $T^E_{\text{QSL}}$ vs. $L$ at three different values of gravitational scalar potential. Under the time bound condition $T^E_{\text{QSL}}/L<1$, we observe quick suppression of entanglement with the increase of gravitational field.

\section{Conclusion}
\label{Sec.6}
We have estimated the QSL time for two-flavor oscillations in the neutrino-antineutrino system in the presence of a gravitational field. It was observed that the QSL time of the initial neutrino flavor state is influenced by the strength of the gravitational field. The investigation of QSL time for the initial flavor state evolution suggests fast-flavor neutrino-antineutrino transitions can occur when the gravitational field strength increases from low to high values. Additionally, we have examined two bipartite entanglement measures: entanglement entropy and capacity of entanglement. The non-zero values of these entanglement measures have revealed that the two-flavor oscillation of the neutrino-antineutrino system exhibits pure bipartite entanglement. Further, we have explored the QSL time for entanglement entropy for the same system. In the proximity of a spinning BH, with the transition from a low to high gravitational field, there was a faster suppression of entanglement during the evolution of a two-flavor neutrino-antineutrino system. However, throughout the work, we assume that neutrinos are moving around a rotating BH at fixed radii. 

\begin{acknowledgement}
The authors would like to acknowledge the project funded by SERB, India, with Ref.No. CRG/2022/003460, for supporting this research.
\end{acknowledgement}

\end{document}